\documentclass[aps,prl,reprint,groupedaddress]{revtex4-2}
\usepackage{graphicx}% Include figure files
\usepackage{url}
\usepackage{color}

\begin{document}

%\preprint{}

\title{Precise measurement of differential cross sections of the $\Sigma^{-}p \to \Lambda n$ reaction in momentum range 470--650 MeV/$c$}



\author{K. Miwa$^1$, J.K. Ahn$^2$, Y. Akazawa$^3$,  T. Aramaki$^1$, S. Ashikaga$^4$, S. Callier$^{5}$, N. Chiga$^{1}$, S. W. Choi$^{2}$, H. Ekawa$^{6}$, \\
P. Evtoukhovitch$^{7}$, N. Fujioka$^{1}$, M. Fujita$^{8}$, T. Gogami$^{4}$, T. Harada$^{4}$, S. Hasegawa$^{8}$, S. H. Hayakawa$^{1}$, R. Honda$^{3}$, \\
S. Hoshino$^{9}$, K. Hosomi$^{8}$, M. Ichikawa$^{4, 14}$, Y. Ichikawa$^{8}$, M. Ieiri$^{3}$, M. Ikeda$^{1}$, K. Imai$^{8}$, Y. Ishikawa$^{1}$, S. Ishimoto$^{3}$, \\
W. S. Jung$^{2}$, S. Kajikawa$^{1}$, H. Kanauchi$^{1}$, H. Kanda$^{10}$, T. Kitaoka$^{1}$, B. M. Kang$^{2}$, H. Kawai$^{11}$, S. H. Kim$^{2}$, \\
K. Kobayashi$^{9}$, T. Koike$^{1}$, K. Matsuda$^{1}$, Y. Matsumoto$^{1}$, S. Nagao$^{1}$, R. Nagatomi$^{9}$, Y. Nakada$^{9}$, M. Nakagawa$^{6}$, \\
I. Nakamura$^{3}$, T. Nanamura$^{4, 8}$, M. Naruki$^{4}$, S. Ozawa$^{1}$, L. Raux$^{5}$, T. G. Rogers$^{1}$, A. Sakaguchi$^{9}$, T. Sakao$^{1}$, \\
H. Sako$^{8}$, S. Sato$^{8}$, T. Shiozaki$^{1}$, K. Shirotori$^{10}$, K. N. Suzuki$^{4}$, S. Suzuki$^{3}$, M. Tabata$^{11}$, C. d. L. Taille$^{5}$, \\
H. Takahashi$^{3}$, T. Takahashi$^{3}$, T. N. Takahashi$^{15}$, H. Tamura$^{1,  8}$, M. Tanaka$^{3}$, K. Tanida$^{8}$, Z. Tsamalaidze$^{7, 12}$, \\
M. Ukai$^{3, 1}$, H. Umetsu$^{1}$, S. Wada$^{1}$, T. O. Yamamoto$^{8}$, J. Yoshida$^{1}$, K. Yoshimura$^{13}$}

\affiliation{$^1$ Department of Physics, Tohoku University, Sendai 980-8578, Japan.}
\affiliation{$^2$ Department of Physics, Korea University, Seoul 02841, Korea}
\affiliation{$^3$ Institute of Particle and Nuclear Studies (IPNS), High Energy Accelerator Research Organization (KEK), Tsukuba 305-0801, Japan}
\affiliation{$^4$ Department of Physics, Kyoto University, Kyoto 606-8502, Japan}
\affiliation{$^{5}$ OMEGA Ecole Polytechnique-CNRS/IN2P3, 3 rue Michel-Ange, 75794 Paris 16, France}
\affiliation{$^{6}$ High Energy Nuclear Physics Laboratory, RIKEN, Wako, 351-0198, Japan}
\affiliation{$^{7}$ Joint Institute for Nuclear Research (JINR), Dubna, Moscow Region 141980, Russia}
\affiliation{$^8$ Advanced Science Research Center (ASRC), Japan Atomic Energy Agency (JAEA), Tokai, Ibaraki 319-1195, Japan}
\affiliation{$^9$ Department of Physics, Osaka University, Toyonaka 560-0043, Japan}
\affiliation{$^{10}$ Research Center for Nuclear Physics (RCNP), Osaka University, Ibaraki 567-0047, Japan}
\affiliation{$^{11}$ Department of Physics, Chiba University, Chiba 263-8522, Japan}
\affiliation{$^{12}$ Georgian Technical University (GTU), Tbilisi, Georgia}
\affiliation{$^{13}$ Department of Physics, Okayama University, Okayama 700-8530, Japan}
\affiliation{$^{14}$ Meson Science Laboratory, Cluster for Pioneering Research, RIKEN, Wako, 351-0198, Japan}
\affiliation{$^{15}$ Nishina Center for Accelerator-based Science, RIKEN, Wako, 351-0198, Japan}

\collaboration{J-PARC E40 Collaboration}

%\input{author.tex}
\date{\today}

\begin{abstract}
The differential cross sections of the $\Sigma^{-}p \to \Lambda n$  reaction were measured accurately for the $\Sigma^{-}$ momentum ($p_{\Sigma}$) ranging from 470 to 650 MeV/$c$ at the J-PARC Hadron Experimental Facility.
Precise angular information about the $\Sigma^{-}p \to \Lambda n$ reaction was obtained for the first time by detecting approximately 100 reaction events at each angular step of $\Delta \cos \theta = 0.1$.
The obtained differential cross sections show slightly forward-peaking structure in the measured momentum regions.
The cross sections integrated for $-0.7 \le \cos \theta \le 1.0$ were obtained as $22.5\pm0.68 \rm{(stat.)} \pm0.65 \rm{(syst.)} $ mb and $15.8\pm0.83 \rm{(stat.)}\pm0.52 \rm{(syst.)}$ mb for $470<p_{\Sigma}({\rm MeV}/c)<550$ and $550<p_{\Sigma}({\rm MeV}/c)<650$, respectively.
These results show a drastic improvement compared to past measurements of the hyperon-proton scattering experiments.
They will play essential roles in updating the theoretical models of the baryon-baryon interactions.
\end{abstract}

%\keywords{}

\maketitle

%\section{}
%\subsection{}
%\subsubsection{}

% If in two-column mode, this environment will change to single-column
% format so that long equations can be displayed. Use
%\begin{widetext}
%\end{widetext}

The interactions between octet baryons, that is, baryon-baryon ($BB$) interactions including hyperon-nucleon ($YN$) interactions are fundamental information for describing nuclear systems including hyperons such as hypernuclei and neutron stars.
Historically, experimental data attributed to a pure two-body $YN$ system is quite limited due to various difficulties involved in conducting hyperon-proton scattering experiments \cite{Sechi-zorn:1968, Alexander:1968, Kadyk:1971, Hauptman:1977, Engelmann:1966, Eisele:1971, Stephen:1970, Kondo:2000, Kanda:2005}.
%Therefore, historically, the $YN$ interactions have been investigated from hypernuclear data because their binding energies and energy levels reflect the $YN$ interactions \cite{Hashimoto:2006, Yamamoto:2010}.
However, there has been recent progress in obtaining the two-body $YN$ interaction from a two-body system.
%The high-energy heavy ion collision experiment such as ALICE and STAR collaborations are one of the best places to investigate two-body $YN$ interactions from the femtoscopy, where particle correlations between many hadron pairs are measured to extract the interactions between the hadrons by assuming the source size at the collision.

We (J-PARC E40 collaboration) reported accurate measurements of the differential cross sections of the $\Sigma^{-}p$ elastic scattering in the momentum range from 470 to 850 MeV/$c$ \cite{Miwa:2021}.
This measurement first provided accurate differential information, which is essential for determining the $P$ and higher-wave interactions.
The CLAS collaboration also reported the updated total cross sections of the $\Lambda p$ elastic scattering for the $\Lambda$ momentum between 0.9 and 2.0 GeV/$c$ \cite{Rowley:2021}.
The ALICE \cite{ALICE:2019, ALICE:LambdaLambda, ALICE:2020, ALICE:PLB802, ALICE:XiN, ALICE:OmegaN, ALICE:LN_SN}, and STAR \cite{STAR:2015, STAR:OmegaP} collaborations measured particle correlations not only for the hyperon-nucleon pairs but also for the hyperon-hyperon pairs.
These measurements, which are sensitive to small values of relative momentum, constitute new experimental methods for determining the $S$-wave interaction \cite{Morita:2015, A_Ohnishi:2016, Morita:2020}.
%Experimentally, the modern methods to study the two-body $YN$ interactions are developed and new data are expected to be provided in future. 

The Nijmegen group \cite{Rijken:1999, Rijken:2010, Nagels:2019} and the J\"{u}lich group \cite{Haidenbauer:2005} developed theories behind $BB$ interactions using a boson-exchange model and considering a broken flavor SU(3) symmetry.
The quark cluster model (QCM) was proposed to explain the origin of the short-range repulsive core in the nucleon-nucleon ($NN$) interactions by considering the effects of the Pauli principle for the quarks and the color magnetic interaction between them\cite{Oka:1986}. 
The Kyoto-Niigata group constructed a realistic description by incorporating an effective meson exchange potential into QCM to represent the middle- and long-range interactions \cite{Fujiwara:2007}.

 $BB$ interactions have also been intensively studied using modern theoretical frameworks, such as  the lattice QCD simulations \cite{Aoki:2012, Inoue:2012, Nemura:2018} and the chiral effective field theory ($\chi$EFT).
% The lattice QCD is a powerful method for deriving the $YN$ potentials from the first principal in QCD especially for multi-strangeness sector such as $\Xi N$ and $\Omega N$ systems \cite{Aoki:2012, Inoue:2012, Nemura:2018, Iritani:2019, Sasaki:2020}.
 Lattice QCD potentials were used to analyze the particle correlations \cite{Iritani:2019, Sasaki:2020}.
 %For the $S=-1$ sector, there are still technical challenges for getting better precision in the lattice QCD simulation.
% However, theoretical improvement are on going using a Misner's method. 
$\chi$EFT is widely used for deriving the $NN$ force because it has an underlying chiral symmetry in QCD and a power counting feature to improve the calculation systematically by moving to a higher order \cite{Epelbaum:2009}.
%Multi baryon forces appear naturally and automatically in a consistent implementation of the $\chi$EFT framework.
$\chi$EFT has been extended to the hyperon sector too \cite{Haidenbauer:2013, Haidenbauer:2020, Song:2021}.

%Two-body $YN$ interactions are fundamental inputs for describing the many-body nuclear system with hyperons, including neutron stars.
%Now, it is highly expected that a realistic $YN$  interaction is constructed by gathering the theoretical and experimental efforts.
Realistic $YN$ interaction models, which should be constructed by gathering the theoretical and experimental efforts,  will create a new trend in hypernuclear  physics.
For example, the no-core shell model calculations based on the $\chi$EFT extended to the $YN$ sector were recently performed to describe the $p$-shell hypernuclei \cite{Wirth:2014, Wirth:2018, Le:2020, Le:2021}. 
A realistic $YN$ interaction is also essential for constructing the equation of state of  neutron stars with microscopic approaches using  bare two-body $YN$  interactions \cite{Burgio:2021}.
 
In this letter, we present new results on the differential cross sections of the $\Sigma^{-}p \to \Lambda n$   reaction in the $\Sigma^{-}$ momentum range 470--650 MeV/$c$ measured in the J-PARC E40 experiment\cite{Miwa:2021, Miwa:2020}.
The $\Sigma p$ scatterings (the $\Sigma^{-}p$ and $\Sigma^{+}p$ elastic scatterings and the $\Sigma^{-} p \to \Lambda n$ reaction) were systematically measured in the experiment.
%The systematic measurements of the $\Sigma p$ reactions are quite important to have the unified description of the $BB$ interactions using the partially broken flavor SU(3) symmetry.
%The differential cross sections of the $\Sigma^{-}p \to \Lambda n$ reaction together with the $\Sigma^{-}p$ elastic scattering \cite{Miwa:2021} are essential input to improve the theoretical models of the $BB$ interactions.
%Such a systematic study is indispensable for improving the theoretical models of $BB$ interactions.

The $\Sigma^{-}p$ channel is closely related with the $\Lambda N$ system because of the $\Lambda N$-$\Sigma N$ coupling \cite{Gibson:1972}.
%The $\Lambda N$-$\Sigma N$ coupling is a dominant source of the attraction in the $\Lambda N$ interaction, where the one pion exchange is forbidden from the isospin conservation.
%In the $\chi$EFT extended to the $YN$ sector, by artificially switching off the $\Lambda N$-$\Sigma N$ coupling, the $\Lambda N$ interaction becomes less attractive or repulsive depending on the strength of the $\Lambda N$-$\Sigma N$ coupling \cite{Haidenbauer:2020}.
%The $\Lambda N$-$\Sigma N$ coupling effect plays the important role to explain the binding energies of  the light hypernuclei  \cite{Hiyama:2002, Nemura:2002, Nogga:2002}.
%Especially, the $\Lambda N$-$\Sigma N$  coupling effect in the $\Lambda N N$ system in the neutron-rich $\Lambda$ hypernuclei is the main topics to be investigated by the hypernuclear spectroscopy \cite{Saha:2005, Akaishi:2000, Sugimura:2014, Honda:2017}.
%
The strength of the $\Lambda N$-$\Sigma N$ coupling has been intensively discussed in relation to the so-called hyperon puzzle in neutron stars \cite{Haidenbauer:2017}.
In the nuclear (neutron) matter, the $\Lambda N$-$\Sigma N$ coupling, which is a dominant source of the attraction in some $\Lambda N$ interactions \cite{Haidenbauer:2017},  can be suppressed as a result of Pauli blocking for the intermediate nucleon state.
For interactions with a sizable  $\Lambda N$-$\Sigma N$ coupling potential such as the one in  $\chi$EFT NLO13 interaction, the $\Lambda N$ interaction becomes more repulsive at higher baryon densities compared to that in $\chi$EFT NLO19 interaction with a moderate coupling potential \cite{Haidenbauer:2020}.
Such a scenario in which the $\Lambda N$ interaction becomes repulsive, together with an additional repulsive $\Lambda NN$  three-body force \cite{Petschauer:2017, Kohno:2018}, is hypothesized to prevent the $\Lambda$ particles from appearing in the neutron stars  and to explain neutron stars with two-solar masses \cite{Gerstung:2020}.
To constrain the strength of the two-body $\Lambda N$-$\Sigma N$  coupling, reactions involving the conversion such as $\Sigma^{-}p \to \Lambda n$ are potentially  important.

%In the $\chi$EFT, two versions, that is, NLO13 and NLO19 are constructed from the different sets of the low energy constants (LEC), resulting in the different $\Lambda N$-$\Sigma N$ coupling potentials.
%The NLO13 has the large $\Lambda N$-$\Sigma N$ coupling, whereas the NLO19 has the moderate size of the coupling.
%In the $\chi$EFT, however, the difference in the NLO13 and NLO19, which  indicate the difference of strength of the $\Lambda N$-$\Sigma N$ coupling ,  does not appear in the differential cross section of the $\Sigma^{-}p \to \Lambda  n$ reaction in the present $\Sigma^{-}$  momentum range.
%Instead, they pointed out that the $\Lambda p \to \Sigma^{0}p$ reaction at the $\Sigma N$ threshold in the $\Lambda$ momentum is very sensitive to the $\Lambda N$-$\Sigma N$ coupling.
%However, the accurate differential cross section of the $\Sigma^{-}p \to \Lambda n$ reaction would be a standard data for many theoretical framework to construct or check the $\Lambda N$-$\Sigma N$ coupling.

%\begin{figure}[t]
%  \centerline{\includegraphics[width=0.45\textwidth]{figure/CATCH_Sigmap2.eps}}
%  \caption{Experimental concept of the $\Sigma p$ scattering experiment and the experimental setup around the LH$_{2}$ target.
%Two successive two-body reactions of  the $\Sigma^{-}$ production ($\pi^{-} p \to K^{+} \Sigma^{-}$) and the $\Sigma^{-} p$ scattering ($\Sigma^{-} p \to \Sigma^{-} p$) are detected.
%The $\Sigma p$ scattering events are detected by the CATCH system which surrounds the LH$_{2}$ target.
%  }
%  \label{K18beamline_spectrometer_KURAMA_w_CATCH}
%\end{figure}

The $\Sigma p$ scattering experiment (J-PARC E40)  was performed at the K1.8 beam line in the J-PARC Hadron Experimental Facility.
%Data of the $\Sigma^{-}p$ channels were taken in Spring 2019.
%J-PARC provides high intensity secondary beams such as pion and kaon produced from 30 GeV proton beam with a spill cycle of 5.2 seconds with a beam duration of 2 seconds in the slow extraction mode.
A 1.33 GeV/$c$ $\pi^{-}$ beam of $2.0 \times10^{7}$/spill was produced from a 30 GeV proton beam with a cycle of 5.2 seconds and a beam duration of 2 seconds.
The experimental concept and the experimental setup are shown in Fig. 1 in \cite{Miwa:2021}.
$\Sigma^{-}$ particles were produced by the $\pi^{-}p \to K^{+}\Sigma^{-}$ reaction in a liquid hydrogen (LH$_{2}$) target, and the produced $\Sigma^{-}$ moving in the LH$_{2}$ target interacted with protons.
The momentum of each $\Sigma^{-}$ was measured 
%\color{red}
\color{black}
with an accuracy of approximately 5 MeV/$c$ 
\color{black}
as the missing momentum calculated from the momenta of the $\pi^{-}$ beam and the outgoing $K^{+}$ analyzed by the K1.8 beam line spectrometer \cite{Takahashi:2012} and the forward magnetic spectrometer (KURAMA), respectively.
In total, 1.62 $\times 10^{7}$ $\Sigma^{-}$ particles  were used to search for the $\Sigma^{-}$-induced secondary reactions.
Secondary reactions such as the $\Sigma^{-}p \to \Lambda n$ reaction  were identified kinematically from the data of the charged particles in the final state using the CATCH system surrounding the LH$_{2}$ target, which comprises a cylindrical scintillation fiber tracker (CFT), a bismuth germanate calorimeter (BGO), and a scintillator hodoscope (PiID) coaxially arranged outwards from the center \cite{Akazawa:2021, Akazawa:2019}.
The tracks of the charged particles were reconstructed using CFT, and their kinetic energies were measured using BGO.
A detailed analysis of the $\Sigma^{-}$ identification and the secondary reaction with CATCH is found in \cite{Miwa:2021}.
In this letter, we focus on the analysis to derive the differential cross sections of the $\Sigma^{-}p \to \Lambda n$ reaction.

%Analysis of the $\Sigma^{-}p \to \Lambda n$ reaction consists of three components.
%First, the momentum-tagged $\Sigma^{-}$ particles are identified from the analysis of the beam-line and KURAMA spectrometers.
%%The $\Sigma^{-}$ production vertex was required to be inside of the LH$_{2}$ vessel.
%In total, the $\Sigma^{-}$ particles of 1.62 $\times 10^{7}$, whose momenta ranged from 470 to 850 MeV/$c$, were identified from the missing mass spectrum of the $\pi^{-}p \to K^{+}X$ reaction.
%Subsequently, the $\Sigma^{-} p \to \Lambda n$ reaction and other secondary reactions are identified from the analysis of CATCH for the $\Sigma^{-}$ production events.
%Finally, the differential cross section is derived.
%The first two components are common to the analysis of the $\Sigma^{-} p$ elastic scattering and are described in detail in \cite{Miwa:2021}.
%In this letter, we focus on the analysis for the derivation of the differential cross section of the $\Sigma^{-}p \to \Lambda n$ reaction.

For the analysis of the $\Sigma^{-}p \to \Lambda n$ reaction in the LH$_{2}$ target, both a proton and a $\pi^{-}$ were required to be detected by CATCH in coincidence with the $\Sigma^{-}$ production
\color{black}
 within a time interval of $\pm$10 ns.
 \color{black}
Particle identification between $\pi^{-}$ and proton was performed by the $dE$-$E$ method between the partial  energy deposit ($dE$) in CFT and the total energy deposit ($E$) in BGO.
One can refer to Fig. 8 in \cite{Miwa:2021}.
The kinetic energy and direction of the proton were measured using CATCH.
However, the thickness of BGO was not sufficient for $\pi^{-}$ to be stopped, and only the direction of the $\pi^{-}$ was obtained by the tracking using CFT.
Therefore, a certain kinematic assumption is necessary to estimate the magnitude of the $\pi^{-}$ momentum.
In the analysis of the $\Sigma^{-}p \to \Lambda n$ reaction, the $\pi^{-}$ momentum was determined such that the invariant mass of the $\pi^{-}$ and proton became the $\Lambda$ mass using the momentum of the proton and the opening angle between the two tracks.
%When both the proton and $\pi^{-}$ hit the same segment of BGO, such events were removed to prevent the miscalculation of the kinetic energy of the proton.
A vertex point defined as the closest point between the two tracks was required to be within 40 mm from the target center in the $xy$ plane, which is perpendicular to the beam axis.
We assume that this vertex point is the decay point ($vtx_{decay}$) of the scattered $\Lambda$.
The closest distance at $vtx_{decay}$ was also required to be less than 5 mm.
%A scattering vertex ($vtx_{scat}$) of the $\Sigma^{-}p \to \Lambda n$ reaction is defined as the closest point between the $\Sigma^{-}$ track estimated from the spectrometer's information and the scattered $\Lambda$ track which was estimated from the reconstructed momentum and the $vtx_{decay}$ information.
%We required conservatively that $vtx_{scat}$ should be within 40 mm from the target center in the $xy$ plane and the closest distance between the two tracks should be less than 20 mm.

\begin{figure}[]
\begin{center}
%\begin{minipage}{6.cm}
% \includegraphics[width=0.98\textwidth]{figure/showSigmaMPScatWithFit4_wBG_result_0_c1.eps}
%\end {minipage}
%\hspace{0.2cm}
%\begin{minipage}{6.cm}
% \includegraphics[width=0.98\textwidth]{figure/showSigmaMPScatWithFit4_wBG_result_0_c2.eps}
% \end{minipage}
 %
% \begin{minipage}{6.cm}
% \includegraphics[width=0.98\textwidth]{figure/showSigmaMPScatWithFit4_wBG_result_0_c3.eps}
%\end {minipage}
%\hspace{0.2cm}
%\begin{minipage}{6.cm}
% \includegraphics[width=0.98\textwidth]{figure/showSigmaMPScatWithFit4_wBG_result_0_c4.eps}
% \end{minipage}
%  \centerline{\includegraphics[width=0.6\textwidth]{figure/showSigmaMPScatWithFit4_wBG_result_0_c5.eps}}
  \centerline{\includegraphics[width=0.55\textwidth]{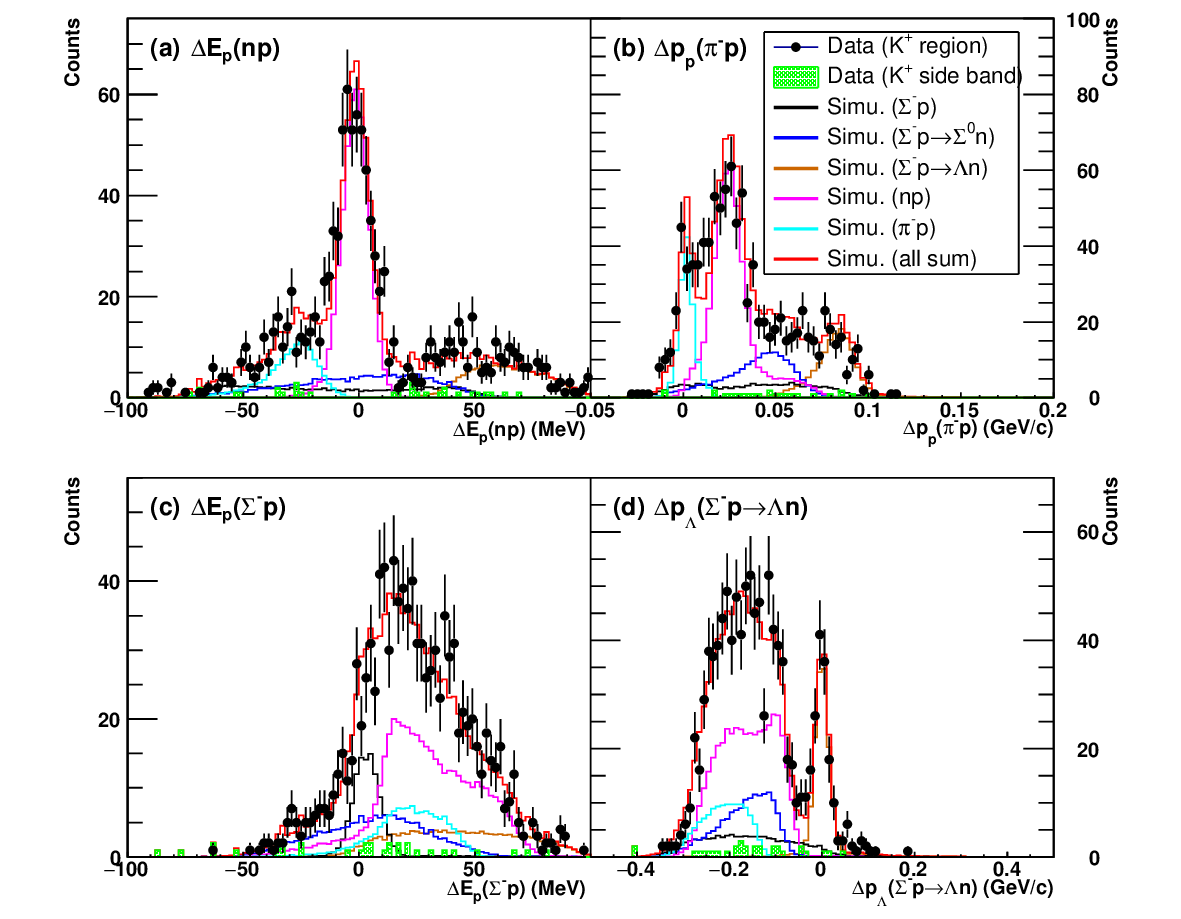}}
   \caption{ Kinematical consistency between the measured energy (momentum) and the calculated one from the measured scattering angle assuming the four different scattering processes,
   (a) $\Delta E_{p} (np)$, 
   (b) $\Delta p_{p} (\pi^{-}p)$,
   (c) $\Delta E _{p}(\Sigma^{-}p)$, and 
   (d) $\Delta p_{\Lambda} (\Sigma^{-}p \to \Lambda n)$ distributions, 
    for the angular region of $0.3 \le \cos \theta \le 0.4$ for the $\Sigma^{-}$ momentum between 470 and 550 MeV/$c$.
Data points with error bars and green-hatched histograms show the experimental data for the $K^{+}$ region and the side-band region of $K^{+}$ in the mass-squared spectrum, respectively.
Simulated spectra for the assumed reactions are also shown, and  the histogram of the red line shows the sum of these spectra.
%The scale factor of each reaction was determined by fitting this four spectra simultaneously.
   }
      \label{showLambdaNConvWithFit4_result_c6}
   \end{center}
\end{figure}

The $\Lambda$ momentum ($p_{\Lambda}^{(\Lambda \to p\pi^{-})}$) reconstructed with the assumption of the $\Lambda \to p \pi^{-}$ decay is checked to determine whether $p_{\Lambda}^{(\Lambda \to p\pi^{-})}$ is consistent with the momentum ($p_{\Lambda}^{(\Sigma^{-}p\to \Lambda n)}$) calculated based on the $\Sigma^{-}p \to \Lambda n$ kinematics from the initial $\Sigma^{-}$ momentum and the $\Sigma^{-}p \to \Lambda n$ scattering angle.
% which is defined from the momentum vectors of the $\Sigma^{-}$ and the $\Lambda$  based on the $\Sigma^{-}p \to \Lambda n$ kinematics. 
We define $\Delta p_{\Lambda}(\Sigma^{-}p \to \Lambda n)$ as the difference between $p_{\Lambda}^{(\Lambda \to p\pi^{-})}$ and $p_{\Lambda}^{(\Sigma^{-}p \to \Lambda n)}$; that is, $\Delta p_{\Lambda} (\Sigma^{-}p \to \Lambda n) = p_{\Lambda}^{(p\pi^{-})} - p_{\Lambda}^{(\Sigma^{-}p \to \Lambda n)}$.
Data points in Fig. \ref{showLambdaNConvWithFit4_result_c6} (d) show the $\Delta p_{\Lambda} (\Sigma^{-}p \to \Lambda n)$ distribution for the 
scattering angle of $0.3 \le \cos \theta \le 0.4$ in the c.m. system for the $\Sigma^{-}$ momentum between 470 and 550 MeV/$c$.
Here, the scattering angle $\theta$ is defined as the angle between the $\Sigma^{-}$ beam and the scattered $\Lambda$.
The peak structure around $\Delta p_{\Lambda} (\Sigma^{-}p \to \Lambda n)=0$ represents the $\Sigma^{-}p \to \Lambda n$ events.

The broad structure in the $\Delta p_{\Lambda} (\Sigma^{-}p \to \Lambda n)$ distribution on the left side of the peak is attributed to other secondary reactions.
As the source of the other secondary reactions, the $\Sigma^{-}p$ elastic scattering and the $\Sigma^{-}p \to \Sigma^{0}n$ reactions are considered.
The scatterings between a target proton and decay products of the $\Sigma^{-} \to n \pi^{-}$ decay, that is, $np$  and $\pi^{-}p$ scatterings, are also taken into account.
%To estimate these contributions, the kinematical consistency was checked for the recoil proton's energy assuming the background kinematics of the $np$, $\pi^{-}p$, and $\Sigma^{-}p$ scatterings.
To identify the source of the background reaction, the measured proton energy was compared with the calculated energies based on the background kinematics.
 For example, in the $np$ scattering case, the energy of the recoil proton was calculated from the initial neutron momentum and the scattering angle between the initial neutron and recoil proton.
In this calculation, the momentum of the initial neutron was obtained by assuming that a $\pi^{-}$ is emitted from the $\Sigma^{-} \to n \pi^{-}$ decay.
Fig. \ref{showLambdaNConvWithFit4_result_c6} (a)  shows the $\Delta E_{p} (np)$ distribution, which is the difference between the measured and calculated kinetic energies of the proton for the $np$ scattering kinematics.
The peak around  $\Delta E_{p}(np) = 0$ corresponds to the $np$ scattering event.
We also define the $\Delta p_{p} (\pi^{-}p)$ (and $\Delta E_{p}(\Sigma^{-}p)$ ) values, representing the difference between the measured momentum (and the measured kinetic energy)  of the proton and the calculated one assuming the $\pi^{-}p$ ( and $\Sigma^{-}p$)  scatterings, respectively, as shown in Figs. \ref{showLambdaNConvWithFit4_result_c6} (b) and (c).
The effect of  misidentification of the initial  $\Sigma^{-}$ particle owing to the contamination of $\pi^{+}$ and protons in the $K^{+}$ selection is also shown as green-hatched histograms in Fig.  \ref{showLambdaNConvWithFit4_result_c6}, obtained by selecting the side-band region of $K^{+}$ in the mass-squared distribution detected by the KURAMA spectrometer \cite{Miwa:2021}.

To estimate the contribution from each secondary reaction, we fit the four $\Delta E$ and $\Delta p$ spectra simultaneously with the simulated spectra of five possible reactions, as shown by the colored spectra in Fig. \ref{showLambdaNConvWithFit4_result_c6}.
Realistic resolutions of the detectors and efficiencies were taken into consideration in the simulation. 
Refer to \cite{Miwa:2021} for a detailed description.
The sum of these spectra reproduces the $\Delta E$ and $\Delta p$ spectra well.
Fortunately, the background reactions are kinematically separated from the $\Sigma^{-}p \to \Lambda n$ reaction, except for the $\Sigma^{-}p$ elastic scattering, as shown by the histogram with a black line in Fig. \ref{showLambdaNConvWithFit4_result_c6} (d).
%However, in the backward angular region of $\cos \theta \le -0.4$, the $\pi^{-}p$ and $np$ background overlaps with the $\Sigma^{-} p \to \Lambda n$ reaction.
%Therefore, the $np$ and $\pi^{-}p$ events are removed by rejecting the peak region of the $\Delta E (np)$ and $\Delta p (\pi^{-}p)$ as shown by arrows in Fig. \ref{showLambdaNConvWithFit4_result_c6} (a) and (b).

The differential cross section is defined as

\begin{equation}
\frac{d \sigma}{d \Omega} = \frac{\sum_{i_{vtz}} \frac{N_{scat}(i_{vtz}, \cos \theta)}{\epsilon(i_{vtz}, \cos \theta)}}{\rho  N_{A}  L  \Delta \Omega}, \label{equ_cs}
\end{equation}
where $\rho$, $N_{A}$, and  $L$ represent the target density, Avogadro's number, and the total flight length of the $\Sigma^{-}$ hyperons in the LH$_{2}$ target, respectively.
%The values of the numerator depend on the vertex position of the $\Sigma^{-}p \to \Lambda n$ reaction.
$i_{vtz}$ represents the index of the $z$ vertex position from $-150$ mm to 150 mm with an interval of 30 mm.
For a scattering angle $\theta$ in the c.m. frame and a $z$ vertex position of $i_{vtz}$, $N_{scat}(i_{vtz}, \cos \theta)$  and $\epsilon(i_{vtz}, \cos \theta)$ represent the number of  $\Sigma^{-}p \to \Lambda n$ reaction events and the detection efficiency of the CATCH system, respectively.
The numerator is the efficiency-corrected number of scattering events.
$\Delta \Omega$ represents the solid angle for each scattering angle.
Regarding the total flight length $L$, the same value as in the $\Sigma^{-}p$ elastic scattering analysis was used \cite{Miwa:2021}.

\begin{figure}[]
  \centerline{\includegraphics[width=0.5\textwidth]{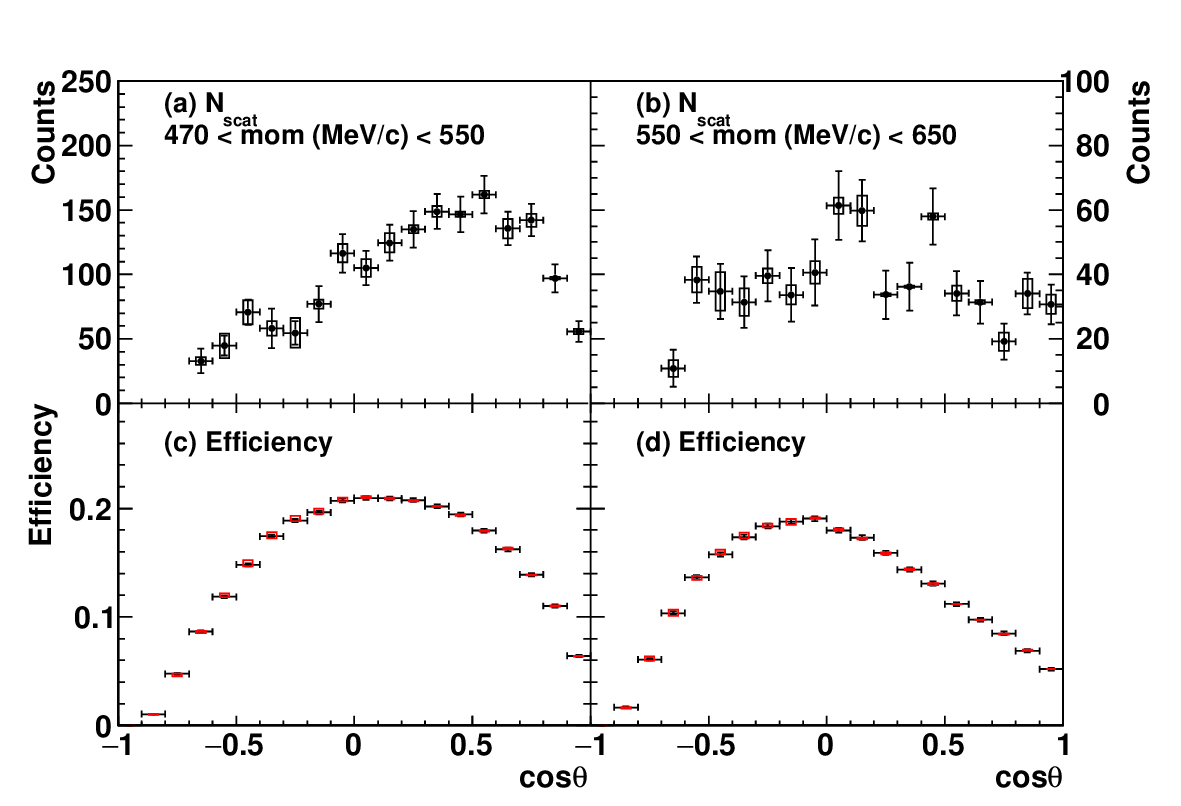}}
  \caption{ 
  (a), (b) Number of  $\Sigma p \to \Lambda n$ reaction events detected for each scattering angle for two $\Sigma^{-}$ beam momentum regions.
\color{black}
  The error bars and boxes show the statistical and systematic errors, respectively.
  \color{black}
  (c), (d) Detection efficiency for the $\Sigma^{-}p \to \Lambda n$ scattering events for two $\Sigma^{-}$ beam momentum regions.
  These efficiencies are the averaged values for all vertex regions of $-150$ $<$ $vtz$ (mm) $<$ 150.
\color{black}
  The red boxes show the systematic uncertainties originating from the tracking efficiency of CFT.
\color{black}  
  }
  \label{compLambdaNEff_Ln_c4}
\end{figure}

\begin{figure*}[]
  \centerline{\includegraphics[width=1.0\textwidth]{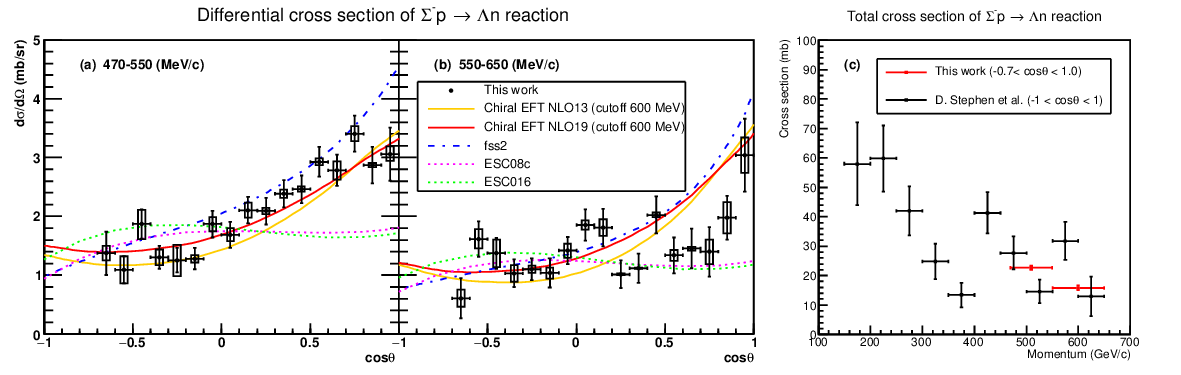}}
  \caption{ (a), (b) Differential cross sections obtained in the present experiment (black points) for two momentum regions.
The error bars and boxes show the statistical and systematic uncertainties, respectively.
The dotted magenta and green lines represent the calculations by the Nijmegen ESC08c \cite{Rijken:2010} and ESC16 \cite{Nagels:2019} interactions based on the boson-exchange model.
The dot-dashed (blue) line shows the calculation using the fss2 model, including QCM \cite{Fujiwara:2007}.
The solid orange and red lines show the calculations using two versions of the extended $\chi$ EFT model, NLO13 \cite{Haidenbauer:2013} and NLO19 \cite{Haidenbauer:2020}, respectively.
In both cases, the cutoff value of 600 MeV is used.
(c) Integrated cross section for $-0.7 \le \cos \theta \le 1$ measured in the present experiment (black points).
Past data measured with a bubble chamber \cite{Stephen:1970} are also shown as blue squares.
  }
  \label{showAllLambdaNConvdSdW3.eps}
\end{figure*}

The detection efficiency for the  $\Sigma^{-}p \to \Lambda n$ scattering events ($\epsilon(i_{vtz}, \cos \theta)$) was studied using a realistic Monte Carlo simulation based on the Geant4 package \cite{Agostinelli:2003}, where the realistic angular resolution, the tracking efficiency of CFT, and the realistic energy resolution for BGO were taken into account \cite{Miwa:2021}.
The generated data of the secondary reactions were analyzed by the same analysis program.
The detection ratio for the $\Sigma^{-}p \to \Lambda n$ reaction was obtained for each scattering angle as its detection efficiency.
Figs. \ref{compLambdaNEff_Ln_c4} (c) and (d)  show the efficiencies averaged for the $z$ vertex region, which is denoted as $\bar{\epsilon}(\cos \theta)$, for the momentum regions 470--550 MeV/$c$ and 550--650 MeV/$c$, respectively.
The branching ratio of the $\Lambda \to p \pi^{-}$ decay is included in the efficiency.
The effect of the systematic uncertainty of the tracking efficiency of CFT, which was estimated from calibration measurements of the $pp$ scattering cross sections \cite{Miwa:2021}, is typically 0.5\% - 3\% as represented by the red box in Figs. \ref{compLambdaNEff_Ln_c4} (c) and (d).
In the backward angle around $\cos \theta = -1$, the kinetic energy of the proton from the $\Lambda$ decay is too small to be detected.
Therefore, the efficiency decreases in the backward angles.
The decrease in the efficiency at the forward angles is due to the decreased  acceptance of CATCH.

The number of scattering events was estimated from the $\Delta p_{\Lambda} (\Sigma^{-}p \to \Lambda n)$ spectra for each scattering angle, as shown in Fig. \ref{showLambdaNConvWithFit4_result_c6} (d).
%The background structure was estimated by fitting the all four $\Delta E$ and $\Delta p$ spectra simultaneously with the sum of the spectra of the assumed secondary reactions.
The sum of the simulated background reactions
%, which were obtained from the simultaneous fit of the our $\Delta E$ and $\Delta p$ spectra, 
was used as the background spectrum.
The efficiency-corrected number of scattering events was estimated in several ways by changing the estimation of the scattering events and by changing the background estimation.
In the $z$ vertex-dependent manner, $N_{scat}(i_{vtz}, \cos \theta)$ was obtained by subtracting the simulated background spectrum from the $\Delta p_{\Lambda} (\Sigma^{-}p \to \Lambda n)$ spectrum in each $z$ vertex region.
%The efficiency-corrected scattering number was obtained by correcting  $N_{scat}(i_{vtz}, \cos \theta)$  with the vertex-dependent CATCH efficiency $\epsilon(i_{vtz}, \cos \theta)$ and .
The efficiency-corrected scattering number was obtained in the form of the numerator of equation (\ref{equ_cs}).
Alternatively, we also estimated the efficiency-corrected scattering number with a modified form  of $\sum_{i_{vtz}}N_{scat}(i_{vtz}, \cos \theta) / \bar{\epsilon}(i_{vtz}, \cos \theta)$.
It implies that  the scattering event number of all $z$ vertex bins was corrected by the averaged efficiency for the $z$ vertex position.
In this method, the scattering event number was obtained from the reproduced spectrum for the $\Sigma^{-}p \to \Lambda n$ reaction (the spectrum with a brown line in Fig. \ref{showLambdaNConvWithFit4_result_c6} (d)) in the $\Delta p (\Sigma^{-}p \to \Lambda n)$ spectrum.
Figs. \ref{compLambdaNEff_Ln_c4} (a) and (b) show the scattering event numbers with the statistical errors for each scattering angle for the two $\Sigma^{-}$ momentum regions.
\color{black}
The detected event numbers are approximately 1700 and  630 in total for the momentum regions 470--550 MeV/$c$ and 550--650 MeV/$c$, respectively.
\color{black}
To estimate the effect of the background estimation, we also derived the efficiency-corrected number of scattering events based on a different background spectrum obtained by fitting the $\Delta p (\Sigma^{-}p \to \Lambda n)$ spectrum alone with the simulated spectra.
The difference in the efficiency-corrected number of scattering events
%, which ranged approximately from a few \% to 20\% depending on background structure, 
was treated as the systematic uncertainty.
\color{black}
The sizes of the systematic uncertainties for each angle are indicated by the  boxes in Figs. \ref{compLambdaNEff_Ln_c4} (a) and (b).
\color{black}

Figs. \ref{showAllLambdaNConvdSdW3.eps} (a) and  (b)  show the measured differential cross sections for the $\Sigma^{-}p \to \Lambda n$ scattering.
These results are obtained from approximately 50 times more scattering events than that in the past experiment \cite{Stephen:1970}.
The statistical and systematic errors are represented as the error bars and boxes, respectively.
\color{black}
The sources of the systematic errors are 
(1) total $\Sigma^{-}$ track length;
(2) estimation of the efficiency-corrected number of scattering events, shown by the boxes in Figs. \ref{compLambdaNEff_Ln_c4} (a) and (b); and
(3) CATCH efficiency,  shown by the red boxes in Figs. \ref{compLambdaNEff_Ln_c4} (c) and (d).
The uncertainty in the total $\Sigma^{-}$ track length is less than 1\% \cite{Miwa:2021}.
\color{black}
The main contribution to the systematic errors comes from the uncertainty of the efficiency-corrected number of scattering events including the systematic error in the background estimation.
These sources were quadratically summed.
%The sources of the systematic errors [total $\Sigma^{-}$ track length, CATCH efficiency (red boxes in Figs. \ref{compLambdaNEff_Ln_c4} (c) and (d)), and estimation of the efficiency-corrected number of scattering events (boxes in Figs. \ref{compLambdaNEff_Ln_c4} (a) and (b))] were quadratically summed.

In these momentum ranges, the differential cross sections of the $\Sigma^{-}p \to \Lambda n$ reaction show a slightly forward peak structure.
In contrast to the $\Sigma^{-}p$ elastic scattering \cite{Miwa:2021}, sizable contributions also exist for the backward angular region.
Figs. \ref{showAllLambdaNConvdSdW3.eps} also shows predictions by various theoretical models.
%This is a contrast to the $\Sigma^{-}p$ elastic scattering.
%In the $\Sigma^{-}p$ elastic scattering in the same momentum region, the differential cross sections show the very forward-peaking structure, whereas the differential cross sections for the backward region ($\cos \theta < -0.3$) are approximately 0.5 mb/sr \cite{Miwa:2021}.
In the momentum region between 470 and 550 MeV/$c$, theoretical calculations by the fss2 including QCM \cite{Fujiwara:2007} and the extended $\chi$EFT model \cite{Haidenbauer:2013, Haidenbauer:2020}, reproduced the measured data adequately.
On the other hand, the Nijmegen models (ESC08c \cite{Rijken:2010} and ESC16 \cite{Nagels:2019}) clearly underestimate the forward angular region.
%The Nijmegen model also underestimate the differential cross section for the $\Sigma^{-}p$ elastic scattering at the forward angular region, especially at the higher momentum region.
In the higher momentum range between 550 and 650 MeV/$c$, the differential cross section becomes flatter  in its angular dependence.
%Although the statistical fluctuations are large in the data, 
Predictions from the fss2  and  the $\chi$EFT seem to overestimate the differential cross section at forward angles.

The theoretical predictions by $\chi$EFT NLO13 and NLO19 are similar, as shown by the red and yellow lines, respectively, in Figs.  \ref{showAllLambdaNConvdSdW3.eps} (a) and (b), even though the strength of the $\Lambda N$-$\Sigma N$ coupling potential is quite different for these two models \cite{Haidenbauer:2020}.
%In the two versions in $\chi$EFT (NLO13 and NLO19), the strengths  of the $\Lambda N$-$\Sigma N$ coupling are different \cite{Haidenbauer:2020}.
%However, the difference in the differential cross section is rather small as shown in red and yellow lines in Fig.  \ref{showAllLambdaNConvdSdW3.eps} (a) (b).
Although the experimental accuracy of the present data is still comparable to the difference between the two models, the present data and  $\Lambda p$ scattering data in future experiments proposed at J-PARC \cite{Miwa:Lp_proposal} will provide new insight into the $\Lambda N$-$\Sigma N$ coupling.
Haidenbauer {\it et al.} also pointed out that a study of the $\Lambda p$ scattering near the $\Sigma N$ threshold would be quite helpful for constraining  the $\Lambda N$-$\Sigma N$ coupling \cite{Haidenbauer:2021}.

The integrated cross sections for $-0.7 \le \cos \theta \le 1.0$ were obtained as $22.5\pm0.68 \rm{(stat.)} \pm0.65 \rm{(syst.)} $ mb and $15.8\pm0.83 \rm{(stat.)}\pm0.52 \rm{(syst.)}$ mb for the momentum regions 470--550 MeV/$c$ and 550--650 MeV/$c$, respectively.
These values were compared with the past measurements \cite{Stephen:1970}, as shown in Fig. \ref{showAllLambdaNConvdSdW3.eps} (c).
%By accumulating approximately 2,500 scattering events, the integrated cross sections were successfully  determined with a few percent statistical errors.
The measured integrated and differential cross sections are invaluable experimental inputs for improving the $BB$ interaction models.
A systematic theoretical investigation of both the $\Sigma^{-}p$ elastic scattering and the $\Sigma^{-}p \to \Lambda n$ reaction can be performed based on these data.

\color{black}
Finally, we compare the present results with the $\Sigma^{-}p$ elastic scattering presented in \cite{Miwa:2021}.
Both results are rather consistent with the predictions by the fss2 and $\chi$EFT.
%, although there exists a sizable difference between data and these theories.
On the other hand, the ESC models underestimate the differential cross sections at the forward angular regions for both channels.
These models should also be compared with the $\Sigma^{+}p$ data, for which analysis is presently ongoing \cite{Nanamura:2021}.
\color{black}

In summary, we successfully measured the differential cross sections of the $\Sigma^{-}p \to \Lambda n$ reaction for the momentum region 470--650 MeV/$c$ at J-PARC.
These results are part of a series of systematic studies of $\Sigma N$ interactions from the two-body $\Sigma^{\pm} p$ scatterings.
The differential cross sections were measured for the wide angular region of $-0.7 \le \cos \theta \le 1.0$ by detecting approximately 100 scattering events for each angular bin of $\Delta \cos \theta = 0.1$.
The total number of the reaction events is approximately 50 times larger than the past experiment.
The differential cross section of the $\Sigma^{-}p \to \Lambda n$ reaction shows a moderate forward-peaking angular distribution.
The integrated cross sections for angular coverage were also obtained with a drastically improved accuracy.
These accurate measurements will play an essential role in establishing realistic $BB$ interaction models.

We would like to thank the staff of the J-PARC accelerator and the Hadron Experimental Facility for their support  in providing the beam during the beam time.
We would also like to thank the staff of CYRIC and ELPH at Tohoku University for their support in providing beams for test experiments for our detectors.
We would like to express our gratitude to  Y. Fujiwara for the theoretical support from an early period of the experimental design and  thank T. A. Rijken and J. Haidenbauer for their theoretical calculations.
We also thank KEKCC and  SINET4.
This work was supported by  the JSPS KAKENHI Grant Numbers 23684011, 15H00838, 15H05442, 15H02079, and 18H03693.
This work was also supported by the Grants-in-Aid Number 24105003 and 18H05403 for Scientific Research from the Ministry of Education, Culture, Science, and Technology (MEXT), Japan.

\bibliography{bibliography}
                      
\end{document}